\documentclass[
    a4paper, 
    aps,
    pra,
    amssymb, 
    amsmath, 
    superscriptaddress, 
    amsfonts, 
    reprint, 
    showkeys, 
    nofootinbib, 
]{revtex4-2}

\usepackage[english]{babel}
\usepackage[utf8]{inputenc}
\usepackage[left=23mm,right=13mm,top=24mm,bottom=25mm,columnsep=15pt]{geometry} 
\usepackage[T1]{fontenc}
\usepackage{mathtools}
\usepackage{physics}
\usepackage[table]{xcolor}
\usepackage{graphicx}
\usepackage{csquotes}
\usepackage{multirow}
\usepackage{rotating} 
\usepackage{booktabs}
\usepackage{adjustbox}
\usepackage{fontawesome}
\usepackage{dcolumn}
\usepackage{dsfont}
\usepackage{placeins}
\usepackage[titletoc,toc,page,title,header]{appendix}
\usepackage[pdftex, pdftitle={Article}, pdfauthor={Author}]{hyperref}
\usepackage{orcidlink}
\usepackage{xspace}
\usepackage{tcolorbox}
\usepackage{datetime}
\usepackage{tabularx}
\usepackage{caption}
\usepackage{CJKutf8}
\tcbuselibrary{listings}
\newdateformat{monthyeardate}{%
  \monthname[\THEMONTH] \THEYEAR
}

\definecolor{codebkg}{HTML}{ebebeb}

\hypersetup{
    colorlinks=true,
    linkcolor=blue,
    filecolor=magenta,      
    urlcolor=blue,
    citecolor=blue
}

\DeclareRobustCommand{\inlinecode}[1]{%
  \tcbox[
    on line,
    boxsep=1pt,
    colback=codebkg,
    colframe=codebkg,
    fontupper=\ttfamily,
    nobeforeafter,
    tcbox raise base,
    size=fbox
  ]{#1}%
}

\begin{document}

\title{
    Qiskit Machine Learning: an open-source library for quantum machine learning tasks at scale on quantum hardware and classical simulators
}

\author{M.~Emre~Sahin\,\orcidlink{0000-0002-5996-0407}}
\affiliation{The Hartree Centre, STFC, Sci-Tech Daresbury, Warrington, WA4 4AD, United Kingdom}

\author{Edoardo~Altamura\,\orcidlink{0000-0001-6973-1897}}
\affiliation{The Hartree Centre, STFC, Sci-Tech Daresbury, Warrington, WA4 4AD, United Kingdom}

\author{Oscar~Wallis\,\orcidlink{0009-0002-7323-2059}}
\affiliation{The Hartree Centre, STFC, Sci-Tech Daresbury, Warrington, WA4 4AD, United Kingdom}

\author{Stephen~P.~Wood\,\orcidlink{0000-0001-7333-2792}}
\affiliation{IBM Quantum, IBM T.J. Watson Research Center, Yorktown Heights, NY 10598, USA}

\author{Anton~Dekusar\,\orcidlink{0000-0001-7335-0667}}
\affiliation{IBM Quantum, IBM Research Europe -- Dublin, Ireland}

\author{Declan~A.~Millar\,\orcidlink{0000-0003-3713-8997}}
\affiliation{IBM Research -- UK}

\author{Takashi~Imamichi\,\orcidlink{0000-0002-4423-6897}}
\affiliation{IBM Quantum, IBM Research -- Tokyo, Tokyo 103-8510, Japan}

\author{Atsushi~Matsuo\,\orcidlink{0000-0003-1071-2696}}
\affiliation{IBM Quantum, IBM Research -- Tokyo, Tokyo 103-8510, Japan}

\author{Stefano~Mensa\,\orcidlink{0000-0002-0938-144X}}
\email{stefano.mensa@stfc.ac.uk}
\affiliation{The Hartree Centre, STFC, Sci-Tech Daresbury, Warrington, WA4 4AD, United Kingdom}

\author{Code contributors}

\date{\today}

\begin{abstract}
    We present Qiskit Machine Learning (ML), a high-level Python library that combines elements of quantum computing with traditional machine learning. The API abstracts Qiskit's primitives to facilitate interactions with classical simulators and quantum hardware. Qiskit ML started as a proof-of-concept code in 2019 and has since been developed to be a modular, intuitive tool for non-specialist users while allowing extensibility and fine-tuning controls for quantum computational scientists and developers. The library is available as a public, open-source tool and is distributed under the Apache version 2.0 license.
\end{abstract}

\maketitle

\section{Introduction}
\label{sec:intro}

The convergence of quantum computing and machine learning promises a prospective shift in both research and industry. Quantum machine learning (QML) leverages the principles of quantum mechanics to potentially enhance or accelerate classical machine learning algorithms, opening new frontiers in fields ranging from materials science to finance. As the field of QML matures, there is a growing need for accessible and powerful software tools that bridge the gap between theoretical QML algorithms and their practical implementation on emerging quantum hardware and simulators.

Qiskit Machine Learning (ML)\footnote{\href{https://github.com/qiskit-community/qiskit-machine-learning}{\textcolor[HTML]{840484}{\faGithub}~{github.com/qiskit-community/qiskit-machine-learning}}}, an open-source module within the Qiskit framework \cite{qiskit2024}, addresses this need by providing a comprehensive and user-friendly platform for exploring the exciting landscape of QML. Built on core Qiskit elements such as primitives, it combines quantum circuit design, simulation, and execution to deliver cutting-edge QML algorithms. Users can experiment with quantum enhancements to established methods, such as quantum kernels for Support Vector Machines, or explore new, fully quantum approaches. Its tight integration with Python and reliance on widely used libraries like NumPy \cite{harris2020array} and scikit-learn \cite{scikit-learn} make it accessible to practitioners in diverse fields, from engineering to the life sciences. It also includes a dedicated API connector to PyTorch \cite{paszke2017automatic} for neural network-based algorithms, seamlessly bridging quantum circuits with modern deep learning frameworks.

Qiskit ML is freely distributed under the Apache 2.0 license, encouraging community participation and open collaboration. Moreover, it sets itself apart from other platforms like PennyLane \cite{bergholm2018pennylane} in its approach to quantum hardware usage. Specifically, Qiskit ML's architecture is deliberately designed to handle quantum hardware workloads, while also allowing experimentation with state-of-the-art classical simulators and models of emulated hardware noise from near-term devices. Moreover, it is designed to be modular and extensible, making the addition of new quantum algorithms or building upon existing ones straightforward. Supported by extensive educational resources and tutorials, Qiskit ML stands at the forefront of QML research, helping students, scientists and developers worldwide investigate the applications of quantum computing for machine learning.

\begin{figure*}[t]
    \centering
    \includegraphics[width=\textwidth, trim={0 65pt 0 0},clip]{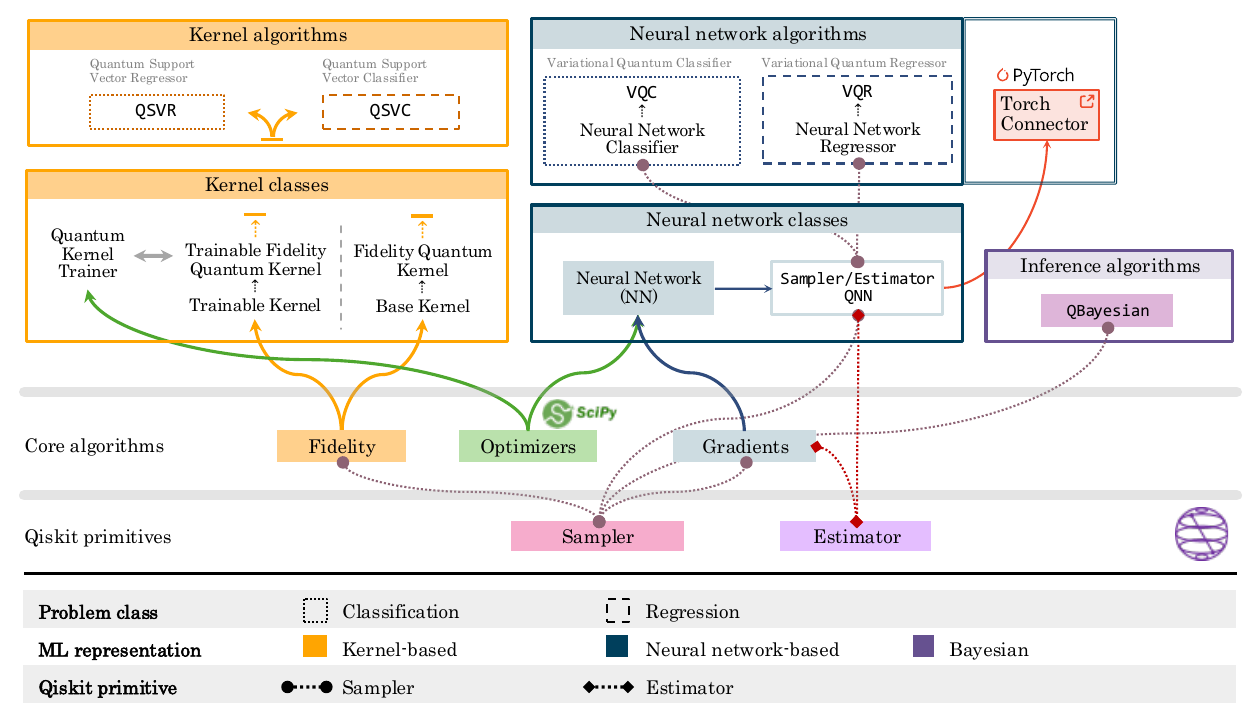}
    \caption{UML diagram of the Qiskit ML library, showing the class hierarchy of the core components, algorithms and dependencies. The diagram categorises machine learning approaches into kernel-based, neural network-based, and Bayesian methods, linking them to their respective problem classes: classification and regression. It highlights key elements such as fidelity quantum kernels, trainable kernels, quantum neural networks, and quantum support vector machines, structured under core algorithms and supported by the Qiskit primitives (\inlinecode{Sampler} and \inlinecode{Estimator}).}
    \label{fig:method-diagram}
\end{figure*}

\section{Design Structure}
\label{sec:structure}
This section outlines the role of Qiskit ML within the broader computing software stack, as illustrated in Fig.~\ref{fig:method-diagram}. Qiskit ML sits at the application level, providing a suite of tools and algorithms that leverage the power of quantum computation for machine learning tasks. It acts as a bridge between high-level machine learning concepts and the underlying quantum hardware or simulators. The design of Qiskit ML prioritises extensibility, modularity and an intuitive user interface, leading to a software package primed for rapid experimentation and prototyping of new and existing QML protocols. The high-level structure includes Quantum Neural Networks (QNNs), Variational Quantum Classifier (VQC) and Regressor (VQR) and Quantum Kernel (QK) Methods for Support Vector Machines, among other QML algorithms. These are provided through workflows shown in the Unified Modelling Language (UML) diagram in Fig.~\ref{fig:method-diagram} and summarised in  Table~\ref{tab:qiskitml-overview}.

\begin{table*}
\centering
\caption{Summary of quantum algorithms-based functions and base models supporting various QML tasks. Gradient estimation techniques, fidelity computations, kernel-based and neural network-based models, and advanced variational and hybrid SVM approaches are outlined for classification, regression, and inference.}
\label{tab:qiskitml-overview}
\begin{tabularx}{\textwidth}{@{}lX@{}}
\bottomrule
\rowcolor{orange!25}
\multicolumn{2}{c}{\textbf{Functions based on Quantum Algorithms}} \\
\toprule
\multirow{3}{*}{\textbf{Gradient Estimation}} 
  & \textbf{Parameter-Shift Rule:} Computes gradients using expectation values of Pauli observables~\cite{schuld2019evaluating}. \\
\cmidrule(lr){2-2}
  & \textbf{Linear Combination of Unitaries:} Approximates gradients by decomposing circuits into unitary operations. \\
\cmidrule(lr){2-2}
  & \textbf{SPSA:} Provides a stochastic estimate of gradients in complex or noisy scenarios~\cite{gacon2021simultaneous}. \\
\midrule
\textbf{Fidelity} 
  & \textbf{ComputeUncompute:} Measures the fidelity between two quantum states by preparing the first state, applying the inverse circuit to the second, and recording the probability of obtaining the all-zeros state. \\
[1ex]
\bottomrule
\rowcolor{blue!50!green!10}
\multicolumn{2}{c}{\textbf{Base Models}} \\
\toprule
\multirow{2}{*}{\textbf{Kernel-based Models}} 
  & \textbf{FidelityQuantumKernel:} Constructs a quantum kernel matrix from a feature map that encodes classical data into quantum states. \\
\cmidrule(lr){2-2}
  & \textbf{TrainableFidelityQuantumKernel:} Extends the fidelity kernel with a parameterised feature map that can be trained for improved performance. \\
\midrule
\multirow{2}{*}{\textbf{Neural Network Based Models}} 
  & \textbf{EstimatorQNN:} Outputs expectation values, suitable for cost function evaluation and feature extraction. \\
\cmidrule(lr){2-2}
  & \textbf{SamplerQNN:} Produces probability distributions, useful for classification and generative tasks. \\
[1ex]
\bottomrule
\rowcolor{orange!25}
\multicolumn{2}{c}{\textbf{Algorithms}} \\
\toprule
\multirow{2}{*}{\textbf{Variational Quantum Algorithms}} 
  & \textbf{VQC/R:} Variational Quantum Classifier/Regressor, uses an optimiser, gradient, and variational circuit to run classification/regression tasks (often used with a sampler). Built on a flexible neural network classifier/regressor. \\
\midrule
\multirow{2}{*}{\textbf{Hybrid SVMs}}
  & \textbf{QSVC/R:} Quantum Support Vector Classifier/Regressor, an extension of scikit-learn's SVC. It uses a Quantum Kernel object to compute fidelity measures and then applies a support vector classifier/regressor on the resulting kernel. \\
\cmidrule(lr){2-2}
  & \textbf{PegasosQSVC:} Extension of QSVC implementing the Pegasos primal sub-gradient solver for the SVM algorithm \cite{shalev2007pegasos}. \\
\midrule
\cellcolor{blue!50!green!10}
\textbf{Inference} 
  & \textbf{QBayesianInference:} Implements Bayesian inference using quantum circuits \cite{low2014quantum, borujeni2021quantum}. \\
\bottomrule
\end{tabularx}
\end{table*}

Embedded within the wider Qiskit ecosystem, Qiskit ML predominantly depends on Qiskit's primitives. It also interfaces with classical machine learning frameworks such as scikit-learn and Python numerical-core libraries like NumPy, enabling a continuous integration of classical and quantum machine learning techniques. Additionally, the models follow SciPy's structural foundation, and there is functionality for integrating neural networks with PyTorch to support the design, training, and inference of hybrid quantum-classical models.

\subsection{Algorithms and Models}
\label{sec:algos}

Qiskit ML integrates the quantum functions and quantum-classical hybrid algorithms (Table~\ref{tab:qiskitml-overview}) described as follows.  \textbf{Gradient estimation} routines compute the gradient $\partial f(\theta)/\partial \theta_i$ of a cost function $f$ parametrized by $\theta$. For example, the Parameter-Shift rule \cite{schuld2019evaluating} calculates exact gradients as:
\begin{equation}
\frac{\partial f(\theta)}{\partial \theta_i} = \frac{f(\theta_i + s) - f(\theta_i - s)}{2 \sin(s)},
\end{equation}
with $s$ as a shift parameter, while the Simultaneous Perturbation Stochastic Approximation (SPSA) \cite{spall1998overview} provides stochastic estimates suitable for near-term quantum hardware \cite{gacon2021simultaneous}.

\textbf{Kernel-based models and SVMs} operate in a feature space to address classification and regression tasks. Fidelity-quantum kernels construct kernel matrices by computing the overlap between quantum states:
\begin{equation}
K(\mathbf{x}, \mathbf{x}') = \left|\bra{\phi(\mathbf{x})}\ket{\phi(\mathbf{x}')}\right|^2,
\end{equation}
where the encoding $\phi$ maps classical data $\mathbf{x}$ to quantum states. Trainable-fidelity kernels introduce learnable parameters to enhance adaptability \cite{gentinetta2023quantum}. Combined with classical support vector machine optimisation (QSVC, QSVR, PegasosQSVC), these methods test quantum-enhanced data separability.

\textbf{Quantum neural networks} utilise learnable parameters for tasks such as regression and classification via the \inlinecode{EstimatorQNN} and \inlinecode{SamplerQNN} classes \cite{Abbas2021}. \textbf{Variational Quantum Algorithms} (VQAs), including VQC and VQR, iteratively optimise quantum circuits through classical routines to minimise a loss function
\begin{equation}
\underset{\theta}{\arg\min}~\mathcal{L}(\theta) = \sum_i \mathcal{C}(y_i, f(\mathbf{x}_i, \theta)),
\end{equation}
with $\mathcal{C}$ as the cost function and $y_i$ the classical labels \cite{schuld2019evaluating}. Finally, \textbf{quantum Bayesian inference} enables probabilistic modelling with quantum circuits under correlated priors \cite{low2014quantum, borujeni2021quantum}.

\subsection{Built-in Qiskit support}
\label{sec:qiskit}

Qiskit ML relies on Qiskit’s primitives to manage quantum operations. At the core, it uses \inlinecode{Sampler} and \inlinecode{Estimator}, which calculate probabilities and expectation values, respectively, for a broad set of use cases in both simulated and real hardware environments. The primitive classes also require using Qiskit's \inlinecode{QuantumCircuit} class and Qiskit observables. This structure allows algorithms and models to be applied across different quantum platforms without changing the high-level logic.

Although not a direct dependency, the IBM Qiskit Runtime\footnote{\href{https://github.com/Qiskit/qiskit-ibm-runtime}{\textcolor[HTML]{840484}{\faGithub}~{github.com/Qiskit/qiskit-ibm-runtime}}} enhances the execution of quantum circuits on IBM Quantum hardware. When using an IBM backend, Qiskit ML leverages Runtime primitives for efficient model training and evaluation on actual quantum devices, exploring the capabilities of near-term quantum computers. Note, when running on hardware, Qiskit ML will often require the inclusion of a transpiler in the form specified by Qiskit \cite{qiskit2024}; this is necessary if the inputted circuit architecture does not adhere to the hardware specifications of topology and native gates. Similarly, Qiskit ML allows the use of Qiskit Aer \cite{qiskit2024} primitives and backends, allowing for fast, parallel simulation of quantum circuits and quantum algorithms on classical high-performance computing facilities equipped with CPUs or GPUs.

\subsection{Classical variational optimizers}
\label{sec:optim}

Qiskit ML supports a variety of optimisation algorithms for refining QML models. The library accommodates optimisers from SciPy, exposed to the user through a uniform interface. Supported optimizers include \inlinecode{L-BFGS-B}, \inlinecode{COBYLA}, and \inlinecode{SLSQP} \cite{byrd1995limited, powell2007view, kraft1988software}, providing a spectrum of gradient-based and gradient-free optimization choices. Furthermore, advanced optimization techniques such as \inlinecode{SPSA}, \inlinecode{ADAM}, and \inlinecode{NFT} are also available within the library \cite{schuld2019evaluating, kingma2014adam, spall1998overview, nakanishi2020sequential}.

\section{Literature using Qiskit ML}
\label{sec:qiskitml-literature}

Since 2019, several research efforts have explored the capabilities of IBM's Qiskit framework in advancing QML applications, particularly for tasks such as classification, regression, and optimisation. Here, we provide an overview of key contributions in this domain, highlighting how Qiskit has been used to implement, test, and optimise QML algorithms. \\

Notably, \citet{havlivcek2019supervised} demonstrated quantum-enhanced feature spaces for supervised learning using variational quantum classifiers and quantum kernel estimators, laying the groundwork for subsequent QML experiments. \citet{glick2024covariant} introduced covariant quantum kernels tailored for data with intrinsic group structure and implemented their approach on a 27‐qubit superconducting processor via Qiskit ML routines. \citet{zoufal2019quantum} demonstrates a quantum generative adversarial network (qGAN) approach that learns and loads random probability distributions into quantum states via a hybrid quantum-classical method implemented using Qiskit ML routines. Moreover, \citet{Abbas2021} compared the effective dimension and generalisation capabilities of quantum neural networks to classical counterparts, demonstrating how Qiskit ML can be leveraged to produce useful metrics. \citet{gentinetta2023quantum} integrates the PEGASOS algorithm into quantum kernel alignment to simultaneously optimise the SVM training and feature-map parameters. \citet{tacchino2019artificial} implement a quantum-like perception architecture on a real quantum processor using Qiskit, demonstrating that quantum-like neurons can efficiently perform nonlinear classification. Several studies have employed Qiskit ML for various tasks. \citet{john2023optimizing} optimised quantum classification algorithms on classical benchmarks. \citet{Jiang2021} implemented a multilayer perceptron for MNIST on IBM quantum processors, while \citet{Santi2023} introduced a hybrid Quantum Text Classifier. Noise resilience has been explored through resilience analysis \cite{Kumar2024} and pulse‐efficient transpilation \cite{Melo2023}, circuit splitting to avoid barren plateaus \cite{tuysuz2023classical}, qGAN training performance under noisy conditions for high-energy physics \cite{borras2023impact}, understanding noise in geometric ML using hardware up to 64 qubits \cite{tuysuz2024symmetry}. Applications include astrophysics with gamma‐ray burst classification from simulated \cite{2025arXiv250117041F} and real data \cite{2008NIMPA.588...52T, 2024arXiv240414133R}, biomedical \cite{demidik2025quantum}, drug discovery \cite{mensa2023quantum}, and efficient kernel alignment through sub-sampling \cite{sahin2024efficient}. 

Particularly, \citet{agliardi2024mitigating} displays the use of Qiskit ML at scale, utilising features such as \inlinecode{FidelityQuantumKernel}, \inlinecode{ComputeUncompute} and the SPSA algorithm. The paper investigates the applicability of fidelity quantum kernels for real-world and synthetic datasets, addressing the challenge of exponential concentration when scaling to large quantum systems. The authors introduce a bit-flip tolerance strategy for error mitigation and demonstrate its effectiveness on noisy hardware with up to 156 qubits. Collectively, these contributions illustrate significant progress in Qiskit-based QML, spanning noise resilience, computational efficiency, and diverse useful applications.
 
These studies collectively illustrate the progress made in Qiskit-based QML, addressing key challenges such as noise resilience, efficiency improvements, and practical applications.

\section{Outlook and Conclusions}
\label{sec:conclusion}

Qiskit ML is an open-source modular framework built on Qiskit that provides a unified, task-oriented interface for QML algorithms. It integrates smoothly with the Python ecosystem used in machine learning and scientific research. The framework supports a wide range of quantum hardware -- from current noisy intermediate-scale devices to future fault-tolerant quantum computers -- allowing researchers to compare various approaches easily and advance QML research and applications. As quantum hardware improves, Qiskit ML is designed to evolve accordingly. Its adaptable, modular building blocks can be customised to tackle more complex machine learning challenges, bridging the gap between current technology and future applications. This adaptability makes it a strong candidate for the platform of choice for applications such as \cite{liu2021rigorous} on early fault-tolerant quantum architectures, ensuring its ongoing relevance for scaling QML workflows on both near-term and next-generation quantum hardware. 

Moreover, the project owners are committed to developing Qiskit ML openly on GitHub under the Apache-2.0 license, offering source code, issue tracking, and pull-request reviews alongside clear guidelines to facilitate contributions from new developers. Active communication channels, including a dedicated \href{https://github.com/qiskit-community/qiskit-machine-learning/issues}{GitHub issue tracker} and the \href{https://qiskit.enterprise.slack.com/archives/C07JE3V55C1}{Slack workspace}, provide direct access to maintainers for rapid defect management and prioritisation of new features driven by community input. This collaborative governance encourages reproducible workflows and maintains Qiskit ML as a dynamic, community-led standard for QML research, adaptable to emerging hardware developments.

\section{Code availability}
Qiskit ML is an open-source library available at \href{https://github.com/qiskit-community/qiskit-machine-learning}{{github.com/qiskit-community/qiskit-machine-learning}} under the Apache 2.0 license. 
\section{Author contributions} 
ES, EA, and OW wrote the manuscript. EA and ES produced the figures and tables. SPW and AD provided technical support in developing the code base. SM was responsible for funding acquisition; ES, EA, and SM were responsible for project administration. All the authors have provided feedback and contributed to the manuscript review. ES, EA, and OW are core developers of Qiskit ML.

\begin{acknowledgments}
    We thank Stefan W\"orner, Christa Zoufal, Voica Radescu, Sam Johnson, and Dhinakaran Vinayagamurthy for support and useful discussions. Moreover, we acknowledge code contributions from those in Appendix~\ref{app:contributors}.
    This work was supported by the Hartree National Centre for Digital Innovation, a UK Government-funded collaboration between STFC and IBM.
    IBM, the IBM logo, and \href{https://www.ibm.com}{www.ibm.com} are trademarks of International Business Machines Corp., registered in many jurisdictions worldwide. Other product and service names might be trademarks of IBM or other companies. The current list of IBM trademarks is available at \href{https://www.ibm.com/legal/copyright-trademark}{www.ibm.com/legal/copytrade}.
\end{acknowledgments}

\newpage
\bibliography{main}

\appendix

\section{Open-source contributors}
\label{app:contributors}

This list includes contributors to the source code of Qiskit ML, and the modules of Qiskit Algorithms that were integrated into our repository. Some of the authors of this manuscript are core developers and also open-source contributors, and do not appear in this list to avoid duplication.

Guillermo Abad L\'opez (\texttt{GuillermoAbadLopez}),
Amira Abbas, (\texttt{amyami187}),
Gabriele Agliardi (\texttt{gabrieleagl}),
Vishnu Ajith (\texttt{charmerDark}),
David Alber (\texttt{david-alber}),
Sashwat Anagolum (\texttt{SashwatAnagolum}),
Eric Arellano (\texttt{Eric-Arellano}),
Luciano Bello (\texttt{1ucian0}),
Martin Beseda (\texttt{MartinBeseda}),
Lev S. Bishop (\texttt{levbishop}),
Prakhar Bhatnagar (\texttt{prakharb10}),
Vasily Bokov (\texttt{VasilyBokov}),
Arnau Casau (\texttt{arnaucasau}),
Isaac Cilia Attard (\texttt{ica574}),
In-ho Choi (\texttt{q-inho}),
Angelo Danducci (\texttt{AngeloDanducci}),
Amol Deshmukh (\texttt{des137}),
Zina Efchary (\texttt{zinaefchary}),
Eli Ebermot (\texttt{ebermot}),
Julien Gacon (\texttt{Cryoris}),
David Galvin,
Gian Gentinetta (\texttt{gentinettagian}),
Frank Harkins (\texttt{frankharkins}),
Shaohan Hu (\texttt{hushaohan}),
Junye Huang (\texttt{HuangJunye}),
Caleb Johnson (\texttt{caleb-johnson}),
Iy\'an M. (\texttt{iyanmv}),
Darsh Kaushik (\texttt{darshkaushik}),
Dariusz Lasecki (\texttt{DariuszLasecki}),
Felix Lehn (\texttt{FelixLehn}),
Jake Lishman (\texttt{jakelishman}),
Sitong Liu (\texttt{sitong1011}),
Manoel Marques (\texttt{manoelmarques}),
Declan A. Millar (\texttt{declanmillar}),
Sanya Nanda (\texttt{SanyaNanda}),
Paul Nation (\texttt{nonhermitian}),
Yuma Nakamura (\texttt{YumaNK}),
Emilio Pel\'aez (\texttt{epelaaez}),
Anna Phan (\texttt{attp}),
Yassh Ramchandani (\texttt{R-Yassh}),
Peter Roeseler (\texttt{proeseler}),
Ruslan Shaydulin (\texttt{rsln-s}),
Francesca Schiavello (\texttt{FrancescaSchiav}),
Su Xiaoleng (\begin{CJK*}{UTF8}{gbsn}苏小冷\end{CJK*}, \texttt{espREssOOOHHH}),
Divyanshu Singh (\texttt{divshacker}),
Iskandar Sitdikov (\texttt{icekhan13}),
Vicente P. Soloviev (\texttt{VicentePerezSoloviev}),
Benjamin Symons (\texttt{Benjamin-Symons}),
Elena Pe\~na Tapia (\texttt{ElePT}),
Soolu Thomas (\texttt{SooluThomas}),
Arne Thomsen (\texttt{Arne-Thomsen}),
Matthew Treinish (\texttt{mtreinish}),
Cenk Tuysuz (\texttt{cnktysz}),
Rishi Nandha Vanchi (\texttt{RishiNandha}),
Nishant Vasan (\texttt{rockywick}),
Ikko Hamamura (\texttt{ikkoham}),
Moritz Willmann (\texttt{MoritzWillmann}),
Matt Wright (\texttt{mattwright99}),
Evgenii Zheltonozhskii (\texttt{Randl}),
Hoang Van Do (\texttt{Vanimiaou}),
Stefan W\"orner (\texttt{stefan-woerner}),
Christa Zoufal (\texttt{Zoufalc}),
\texttt{Desiree},
\texttt{LeCarton},
\texttt{andrenmelo},
\texttt{beichensinn},
\texttt{chetmurthy},
\texttt{rogue-infinity},
\texttt{omarcostahamido},
\texttt{pybeaudouin},
\texttt{rht},
\texttt{sternparky}.

\end{document}